\documentclass[11pt]{article}

\usepackage[T1]{fontenc}
\usepackage[sc]{mathpazo}
\usepackage{amsmath}
\usepackage{amssymb}
\usepackage{enumerate}
\usepackage{amsthm}

\newtheorem{thm}{Theorem}[section]

\usepackage{amsfonts,mathrsfs}
\usepackage[style]{fncychap}
\usepackage{graphicx} 
\usepackage{geometry} 
\usepackage{eepic}
\usepackage{ifthen}
\newboolean{ElectronicVersion}
\setboolean{ElectronicVersion}{true} 

\usepackage{hyperref}
\hypersetup{pdfpagemode=UseNone}

\def\be{\begin{equation}}
\def\ee{\end{equation}}
\def\ben{\begin{eqnarray}}
\def\een{\end{eqnarray}}
\def\bea{\begin{eqnarray*}}
\def\bea{\begin{eqnarray*}}
\def\eea{\end{eqnarray*}}

\newenvironment{mylist}[1]{\begin{list}{}{
    \setlength{\leftmargin}{#1}
    \setlength{\rightmargin}{0mm}
    \setlength{\labelsep}{2mm}
    \setlength{\labelwidth}{8mm}
    \setlength{\itemsep}{0mm}}}
    {\end{list}}

\newtheorem{thrm}{Theorem}[section]
\newtheorem{lem}[thrm]{Lemma}
\newtheorem{prop}[thrm]{Proposition}
\newtheorem{cor}[thrm]{Corollary}
\theoremstyle{definition}
\newtheorem{definition}[thrm]{Definition}
\newtheorem{remark}{Remark}[section]

\numberwithin{equation}{section}
\newtheorem{conjecture}{Conjecture}
\newtheorem{unclear}{Unclear}

\def\bpz{\begin{prop}}
\def\epz{\end{prop}}
\def\bcj{\begin{conjecture}}
\def\ecj{\end{conjecture}}
\def\bd{\begin{definition}}
\def\ed{\end{definition}}
\def\bu{\begin{unclear}}
\def\eu{\end{unclear}}
\def\s{\sigma}
\def\bpf{\begin{proof}}
\def\epf{\end{proof}}
\def\blm{\begin{lem}}
\def\elm{\end{lem}}

\newcounter{questionnumber}

\begin{document}
\title{{\Large The geometrical properties of parity and time
reversal operators in two dimensional spaces}
\thanks{This  project is supported by Research Fund, Kumoh National Institute of Technology.}}

\author{Minyi Huang $^{1}$,\, Yu Yang $^{2}$,\,  Junde Wu $^{1}$,\, Minhyung Cho $^{3}$
\footnote{Corresponding author. E-mail:
mignon@kumoh.ac.kr}\\
{\small \it $^{1}$ School of Mathematical Science, Zhejiang University, Hangzhou 310027, People's Republic of China}\\
{\small \it $^{2}$ Department of Mathematics,  National University of Singapore, Singapore 119076, Republic of Singapore}
\\{\small \it $^{3}$ Department of Applied Mathematics, Kumoh National Institute of Technology, Kyungbuk,
730-701, Korea}}
\date{}
\maketitle \mbox{}\hrule\mbox\\

\begin{abstract}
The parity operator $\cal P$ and time
reversal operator $\cal T$ are two important operators in the quantum theory, in particular, in the $\cal PT$-symmetric
quantum theory. By using the concrete forms of $\cal P$ and $\cal T$, we discuss their geometrical properties in two dimensional spaces. It is showed that if $\cal T$ is given, then all $\cal P$ links with the quadric surfaces; if $\cal P$ is given, then all $\cal T$ links with the quadric curves. Moreover, we give out the generalized unbroken $\cal PT$-symmetric condition of an operator. The unbroken $\cal PT$-symmetry of a Hermitian operator is also showed in this way.

\end{abstract}
\mbox{}\hrule\mbox\\

\section{Introduction}

Quantum theory is one of the most important theories in physics. It is a fundamental axiom in quantum mechanics that
the Hamiltonians should be Hermitian, which implies that the values of
energy are real numbers. However, non-Hermitian Hamiltonians are also studied in physics.
One of the attempts is Bender's $\cal PT$-symmetric theory \cite{bender1998real}. In this theory, Bender and his colleagues attributed the reality of the energies to the $\cal PT$-symmetric property,
where $\cal P$ is a parity operator and $\cal T$ is a time
reversal operator. Since then, many physicists discussed the properties of $\cal PT$-symmetric quantum systems  \cite{bender2007making}.
It also has theoretical applications in quantum optics, quantum statistics and quantum field theory \cite{ruter2010observation,chang2014parity,deffner2015jarzynski,jones2010non}. Recently, Bender, Brody and Muller constructed a Hamiltonian operator $H$ with the property that if its eigenfunctions obey a suitable boundary condition, then the associated eigenvalues correspond to the nontrivial zeros of the Riemann zeta function, where $H$ is not Hermitian in the conventional sense, while $iH$ has a broken $\cal PT$-symmetry. This result may shed light on the new application of $\cal PT$-symmetric theory in discussing the Riemann hypothesis \cite{bender2017zeros}. It was discovered by Mostafazadeh that the $\cal PT$-symmetric case can be generalized to a more general pseudo-Hermitian quantum theory, and the generalized $\cal PT$- symmetry was also discussed \cite{Mostafazadeh2010Pseudo, deng2012general}. Smith studied the time reversal operator $\cal T$ satisfying that ${\cal T}^2 = -I$ and the corresponding $\cal PT$-symmetric quantum theory \cite{jones2010non}.

In this paper, by using the concrete forms of $\cal P$ and $\cal T$ in two dimensional spaces, we discuss their geometry properties. It is showed that if $\cal T$ is given, then all $\cal P$ links with the quadric surfaces; if $\cal P$ is given, then all $\cal T$ links with the quadric curves. Moreover, we give out the generalized unbroken $\cal PT$-symmetric condition of an operator $H$. The unbroken $\cal PT$-symmetry of a
Hermitian operator is also showed in this way.

\section{Preliminaries}
In this paper, we only consider two dimensional complex Hilbert space $\mathbb{C}^2$.
Let $L({\mathbb C}^2)$ be the complex vector space of all linear operators on $\mathbb{C}^2$, $I$ be the identity operator on $\mathbb{C}^2$,  $\overline{z}$ be the complex conjugation of complex number $z$.

An operator $\cal T$ on $\mathbb{C}^2$ is said
to be anti-linear if ${\cal T}(sx_{1}+tx_{2})=\overline{s}{\cal
T}(x_{1})+\overline{t}{\cal T}(x_2)$. It is obvious that the composition of two anti-linear operators is a linear operator and the composition of an anti-linear operator and a linear operator is still anti-linear. Similar to linear operators, anti-linear operators can also correspond to a matrix with slightly different laws of operation \cite{Uhlmann2016anti}.


%
%

 A time reversal operator $\cal T$ is an
anti-linear operator which satisfies ${\cal T}^2=I$ or ${\cal T}^2=-I$.
A parity operator $\cal P$ is a linear operator which satisfies
$\mathcal{P}^2=I$ \cite{jones2010non,Mostafazadeh2010Pseudo,deng2012general,wigner2012group}.

The Pauli operators will be used frequently in our discussions.
Given the basis $\{e_i\}_{i=1}^{2}$ of $\mathbb{C}^2$, they are usually defined as follows \cite{Greiner2011quantum}:
\ben
&&\sigma_1(x_1e_1+x_2e_2)=x_2e_1+x_1e_2,\label{P1}\\
&&\sigma_2(x_1e_1+x_2e_2)=-ix_2e_1+ix_1e_2,\label{P2}\\
&&\sigma_3(x_1e_1+x_2e_2)=x_1e_1-x_2e_2.\label{P3}
\een
To put it another way, the representation matrices of $\sigma_1, \sigma_2$ and $\sigma_3$ are:
\begin{equation*}
\begin{pmatrix}0&1 \\ 1&0 \end{pmatrix},
\begin{pmatrix}0&-i \\ i&0 \end{pmatrix},
\begin{pmatrix}1&0 \\ 0&-1 \end{pmatrix}.
\end{equation*}

\noindent Pauli operators have the following useful properties \cite{Greiner2011quantum}:
\begin{eqnarray}
&& \sigma_{i}\sigma_{j}=-\sigma_{j}\sigma_{i}=i\epsilon_{ijk}\sigma_{k},  \quad i\neq j,\label{P4}\\
&& \sigma_{i}^2=I,\label{P5}
\end{eqnarray}
where $i,j,k\in\{1,2,3\}$, $\epsilon_{ijk}$ is the Levi-Civita
symbol:
\[
\epsilon_{ijk}=\left\{
\begin{array}{lr}
\epsilon_{123}=\epsilon_{231}=\epsilon_{312}=1,\\
\epsilon_{132}=\epsilon_{213}=\epsilon_{321}=-1,\\
0, otherwise.
\end{array}
\right.
\]
The well known commutation and anti-commutation relations are:
\begin{eqnarray*}
&& \sigma_{i}\sigma_{j}-\sigma_{j}\sigma_{i}=2i\epsilon_{ijk}\sigma_{k},\\
&& \sigma_{i}\sigma_{j}+\sigma_{j}\sigma_{i}=2\delta_{ij}I,
\end{eqnarray*}
where $i,j,k\in\{1 , 2 , 3\}$ and $\delta_{ij}$ is the Kronecker
symbol.

Denote $I$ by $\s_0$, then $\{\s_0, \s_1, \s_2, \s_3\}$ is a
basis of $L({\mathbb C}^2)$. Moreover, an operator
$\s=t\s_0+x\s_1+y\s_2+z\s_3\in L({\mathbb C}^2)$ is Hermitian if and only if the coefficients
$\{t,x,y,z\}$ are real numbers.

Given the basis $\{e_i\}_{i=1}^{2}$ of $\mathbb{C}^2$ and any vector $x=\sum x_ie_i$,
one can define an important anti-linear operator, namely the conjugation operator ${\cal T}_0$, by ${\cal T}_0(x)=\sum \overline{x_i}e_i$.

Similar to ${\cal T}_0$, one can define another important anti-linear operator $\tau_0$ by
\be\tau_0(x_1e_1+x_2e_2)=-\overline{x_2}e_1+\overline{x_1}e_2.\label{tau0}\ee
Furthermore, define $\tau_1=\tau_0\sigma_1, \tau_2=\tau_0\sigma_2,
\tau_3=\tau_0\sigma_3$, that is, $\tau_i$ is defined to be the composition of $\tau_0$ and $\sigma_i$. The anti-linear operators
$\{\tau_{0}, \tau_{1}, \tau_{2}, \tau_{3}\}$ forms a basis of the
anti-linear operator space of $\mathbb{C}^2$. This basis has the following properties \cite{Uhlmann2016anti}:
\begin{eqnarray*}
&&\tau_0^2=-I,\\
&&\tau_{0}\sigma_{i}=-\sigma_{i}\tau_{0}=\tau_{i},\\
&&\tau_{i}\tau_{0}=-\tau_{0}\tau_{i}=\sigma_{i},\\
&&\tau_{i}\tau_{j}=\sigma_{i}\sigma_{j}=i\epsilon_{ijk}\sigma_{k}\quad(i\neq j),\\
&&\tau_{i}\tau_{j}-\tau_{j}\tau_{i}=2i\epsilon_{ijk}\sigma_{k},
\end{eqnarray*}
where $i,j\in\{1,2,3\}$.

All the equations above can be verified by the using the definitions of Pauli operators and $\tau_0$.
However, for further use, we show that $\tau_{0}\sigma_{i}=-\sigma_{i}\tau_{0}=\tau_{i}$ in detail.
Consider $\tau_2=\tau_0\sigma_2$. By
(\ref{P2}) and (\ref{tau0}), we have
\begin{eqnarray*}
&&\tau_0\sigma_2(x_1e_1+x_2e_2)=i\overline{x_1}e_1+i\overline{x_2}e_2,\\
&&\sigma_2\tau_0(x_1e_1+x_2e_2)=-i\overline{x_1}e_1-i\overline{x_2}e_2.
\end{eqnarray*}
Thus, $\tau_{0}\sigma_{2}=-\sigma_{2}\tau_{0}=\tau_{2}$. Along similar lines, one can verify that $\tau_{0}\sigma_{i}=-\sigma_{i}\tau_{0}=\tau_{i}$ is also valid for $\sigma_1$ and $\sigma_3$.

Moreover, it follows from $\tau_{0}\sigma_{i}=-\sigma_{i}\tau_{0}=\tau_{i}$ that $\sigma_{j}\tau_{i}=\sigma_{j}\tau_0\sigma_i=-\tau_0\sigma_{j}\sigma_i$.
Combining with (\ref{P4}) and (\ref{P5}),
one can further obtain the following relations:
\begin{eqnarray}
&&\sigma_{j}\tau_{i}=\tau_{i}\sigma_{j}=-i\epsilon_{ijk}\tau_{k},\quad i\neq j,\label{ts1}\\
&&\tau_{i}\sigma_{i}=-\sigma_{i}\tau_{i}=\tau_0,\label{ts2}
\end{eqnarray}
where $i,j,k\in\{1,2,3\}$.

With the help of $\{\sigma_i\}$ and $\{\tau_i\}$, ones can determine the concrete forms of $\cal P$ and $\cal T$:
\blm
\label{le:forms}
Let $\cal P$ be a parity operator  and $\cal T$ be a time reversal operator on $\mathbb{C}^2$. Then

(i). Either ${\cal P}=\pm I$ or ${\cal P}=\displaystyle\sum_{i=1}^3a_{i}\sigma_i$, where
 $a_i$ satisfying $\displaystyle\sum_{i=1}^{3}a_{i}^2=1$.
 The latter case is referred to as the nontrivial ${\cal P}$. A nontrivial $\cal P$ has the following matrix representation:
\begin{equation}
P=\begin{pmatrix}a_3&a_1 - ia_2 \\a_1 + ia_2&-a_3\end{pmatrix}\label{P}.
\end{equation}

(ii). ${\cal T}=\epsilon\displaystyle\sum^{3}_{i=0}c_i\tau_i$, where
 $c_i$ are real numbers, if ${\cal T}^2=I$, then $c_1^2+c_2^2+c_3^2-c_0^2=1$; if ${\cal T}^2=-I$, then $c_1^2+c_2^2+c_3^2-c_0^2=-1$, $\epsilon$ is a unimodular complex number \cite{Uhlmann2016anti}.

\elm

\bpf (i). Suppose ${\cal P}=\displaystyle\sum_{i=0}^3a_{i}\sigma_i$. According to the properties of Pauli operators, we have $I={\cal P}^2=(\displaystyle\sum_{i=0}^{3}a_{i}^2)I+2a_{0}(a_1\sigma_1+a_2\sigma_2+a_3\sigma_3)$.
Note that $\{\s_0, \s_1, \s_2, \s_3\}$ is a basis of $L(\mathbb{C}^2)$, we conclude that $\displaystyle\sum_{i=0}^{3}a_{i}^2=1$ and $a_{0}a_{1}=a_{0}a_{2}=a_{0}a_{3}=0$. If $a_{0}\neq 0$, then $a_1=a_2=a_3=0$, which implies that ${\cal P}=\pm I$. If $ a_0=0$, then the only constraint is $\displaystyle\sum_{i=1}^{3}a_{i}^2=1$ and the matrix takes the form in (\ref{P}).

(ii). The proof can be found in \cite{Uhlmann2016anti}.
\epf

\noindent {\bf Example 1}. In (\ref{P}), if we take $a_2=0$, $a_1,
a_3$ are real numbers satisfying that $a_1^2+a_3^2=1$, and denote $a_1$ by
$\sin\alpha$, $a_3$ by $\cos\alpha$, then $\cal P$ has the matrix representation $\left(\begin{array}{*{2}{c@{\;\;}}c}
\cos\alpha & \sin\alpha  \\
\sin\alpha & -\cos\alpha \\
\end{array} \right)=\begin{pmatrix}\cos\alpha & -\sin\alpha  \\
\sin\alpha & \cos\alpha \end{pmatrix}\begin{pmatrix}1 & 0  \\
0 & -1\end{pmatrix}.$ Thus $\cal P$ is composed of a reflection and a rotation.

\noindent {\bf Example 2}. In (\ref{P}), if $a_1=a_2=0$, $a_3=1$, then
$P=\left(\begin{array}{*{2}{c@{\;\;}}c}
1 & 0  \\
0 & -1  \\
\end{array} \right)$. If $a_2=a_3=0$, $a_1=1$, then $P=\left(\begin{array}{*{2}{c@{\;\;}}c}
0 & 1  \\
1 & 0  \\
\end{array} \right)$. These two parity operators were widely used in \cite{bender2007making}.

\section{The existence of  $\cal P$ commuting with $\cal T$}

In physics, it is demanded that ${\cal P}$ and $\cal T$ are commutative, that is, ${\cal PT}={\cal TP}$. In finite dimensional spaces case, by using the  canonical forms of matrices, one can show that if ${\cal T}^2=I$, then such $\cal P$ always exists. In two dimensional case, we can prove it by utilizing Pauli operators.

\begin{thm}\label{thm1} For each time reversal operator $\cal T$, if ${\cal T}^{2}=I$, then there exists a nontrivial parity operator $\cal P$ such that ${\cal PT}={\cal TP}$. If ${\cal T}^{2}=-I$, then there is no $\cal P$ commuting with
$\cal T$ except ${\cal P}=\pm I$.
\end{thm}
\begin{proof}

We will use the following well known equation frequently,
\be
(\sigma\cdot A)(\sigma\cdot B)=(A\cdot B)I+i\sigma\cdot(A\times B),\label{pc}
\ee
where $A$ and $B$ are two vectors in $\mathbb{C}^3$ and $\sigma=(\sigma_1,\sigma_2,\sigma_3)$. The symbols $\cdot$ and $\times$ represent the dot and cross product of vectors, respectively.

(i). When ${\cal T}^{2}=I$.

Let ${\cal T}=\epsilon\displaystyle\sum^{3}_{i=0}c_i\tau_i$ and ${\cal P}=\displaystyle\sum_{i=1}^3a_{i}\sigma_i$, as was given in Lemma \ref{le:forms}.

 According to (\ref{ts1}) and (\ref{ts2}), ${\cal TP}={\cal PT}$ is equivalent to
\bea
(-c_0\sigma_0+\displaystyle\sum_{j=1}^3 c_j\sigma_j)(\displaystyle\sum_{i=1}^3 \overline{a_i}\sigma_i)\tau_0=
(\displaystyle\sum_{i=1}^3 a_i\sigma_i)(c_0\sigma_0-\displaystyle\sum_{j=1}^3 c_j\sigma_j)\tau_0.
\eea
%

Denote $f_i=Re(a_i)$, $b_i=Im(a_i)$, $\tilde{f}=(f_1, f_2, f_3)$,
$\tilde{b}=(b_1, b_2, b_3)$ and $\tilde{c}=(c_1, c_2, c_3)$. Utilizing (\ref{pc}) to expand the equation above, we have
%
\be
(\tilde{f}\cdot\tilde{c})\sigma_0
-\sigma\cdot[\tilde{b}\times\tilde{c}+c_0\tilde{f}]=0.
\ee
It follows that ${\cal TP}={\cal PT}$ is equivalent to
\begin{eqnarray}
&&c_0\tilde{f}+\tilde{b}\times \tilde{c}=0,\label{c1}\\
&&\tilde{f}\cdot\tilde{c}=0\label{c2}.
\end{eqnarray}
Similarly, by utilizing (\ref{pc}) and Lemma \ref{le:forms}, the contraints ${\cal P}^2=I$ and ${\cal T}^2=I$ can be reduced to the equations as follows,
\begin{eqnarray}
&&\tilde{f}\cdot\tilde{b}=0,\label{c3}\\
&&\|\tilde{f}\|^2-\|\tilde{b}\|^2=1,\label{c4}\\
&&\|\tilde{c}\|^2-c_0^2=1.\label{c5}
\end{eqnarray}

Thus, the problem of finding a parity operator $\cal P$ commuting with $\cal T$ reduces to finding the vectors $\tilde{f}$ and $\tilde{b}$ satisfying
$(\ref{c1})-(\ref{c4})$.

 If $c_0=0$, then we can choose $\tilde{b}=0$ and a unit vector $\tilde{f}$ orthogonal to $\tilde{c}$.
Thus all the conditions $(\ref{c1})-(\ref{c4})$ are satisfied.

 If $c_0\neq 0$. Let $\tilde{b}$ be a vector such that $\tilde{b}$ is orthogonal to $\tilde{c}$ and $\|\tilde{b}\|=|c_0|$.
 Moreover, take $\tilde{f}=\frac{1}{c_0}(\tilde{c}\times\tilde{b})$.
 Direct calculations show that such vectors $\tilde{f}$ and $\tilde{b}$ satisfy $(\ref{c1})-(\ref{c4})$, which completes the proof of the existence of $\cal P$.


(ii). When ${\cal T}^{2}=-I$.

The equation (\ref{c5}) is replaced by the following:
 \be
 \|\tilde{c}\|^2-c_0^2=-1.\label{c6}
 \ee

 Thus $c_0\neq 0$. On the other hand, it follows from (\ref{c1}) that \be\tilde{f}=\frac{1}{c_0}(\tilde{c}\times\tilde{b}).\label{tilf}\ee

  Substituting (\ref{c6}) and (\ref{tilf}) into (\ref{c4}), we have $\|\tilde{f}\|^2-\|\tilde{b}\|^2=1<-\frac{1}{c_0^2}\|\tilde{b}\|^2$, which is a contradiction. Thus, when ${\cal T}^{2}=-I$, there is no $\cal P$ commuting with
$\cal T$ except ${\cal P}=\pm I$.

\end{proof}


\begin{remark} When the space is $\mathbb {C}^4$, although ${\cal T}^{2}=-I$, one can find nontrivial $\cal P$ commuting with $\cal T$
\cite{jones2010non}.
\end{remark}

\section{The geometrical properties of $\cal P$ and $\cal T$}

\begin{thm} Let $\cal
T$ be a time reversal operator satisfying ${\cal T}^2=I$. The set of parity operators $\cal P$ commuting with $\cal T$ correspond uniquely to a hyperboloid in $\mathbb{R}^3$.
\end{thm}

\begin{proof}
As was mentioned above, the determination of $\cal P$ is equivalent to finding out $\tilde{f}$ and $\tilde{b}$ satisfying $(\ref{c1})-(\ref{c4})$.
Now consider $\tilde{m}=\tilde{f}+\tilde{b}$.
We shall prove that all the $\tilde{m}$ form a hyperboloid.

To this end, construct a new coordinate system by taking the direction of $\tilde{c}$ as that of the $X'$ axis. The $Y'-Z'$ plane is perpendicular to $\tilde{c}$ and contains the origin point of
$\mathbb{R}^3$. Assume that $\tilde{m}=(x',y',z')$ in the new $X'Y'Z'$ coordinate system.

(i). If $c_0=0$, then it follows from $(\ref{c1})-(\ref{c3})$ that
$\tilde{b}$ is proportional to $\tilde{c}$ and that $\tilde{f}$ is
orthogonal to both $\tilde{c}$ and $\tilde{b}$. Thus, in the new $X'Y'Z'$ coordinate system,
\begin{eqnarray*}
&&\tilde{b}=(x',0,0),\\
&&\tilde{f}=(0,y',z').
\end{eqnarray*}
On the other hand, equation (\ref{c4}), namely $\|\tilde{f}\|^2-\|\tilde{b}\|^2=1$, implies that
\begin{equation}
y'^2+z'^2-x'^2=1\label{af1}.
\end{equation}

It is apparent that one pair of $\tilde{f}$ and $\tilde{b}$ correspond to one point $\tilde{m}=(x',y',z')$, and vice versa.
In addition, $(\ref{af1})$ represents a hyperboloid in $\mathbb{R}^3$.

(ii). If $c_0\neq 0$, then it follows from (\ref{c5}) that $\tilde{c}=(\sqrt{1+c_0^2},0,0)$ in the $X'Y'Z'$ coordinate system.
In addition, suppose $\tilde{b}=(x_0, y_0, z_0)$ in the $X'Y'Z'$ coordinate system.
By (\ref{tilf}), we have
$\tilde{f}=\frac{1}{c_0}(\tilde{c}\times\tilde{b})=\frac{\sqrt{1+c_0^2}}{c_0}(0, -z_0, y_0)$.
Substituting $\tilde{b}$ and $\tilde{f}$ into (\ref{c4}), we have
\begin{equation}
\frac{1}{c_0^2}(y_0^2+z_0^2)-x_0^2=1\label{c_00}.
\end{equation}
Note that $x_0=x',y_0=\frac{\lambda
z'+y'}{1+\lambda^2},z_0=\frac{z'-\lambda y'}{1+\lambda^2}$, where
$\lambda=\frac{\sqrt{1+c_0^2}}{c_0}$. Thus, one pair of $\tilde{f}$ and $\tilde{b}$ correspond to one point $\tilde{m}=(x',y',z')$, and vice versa. Moreover, it follows from (\ref{c_00}) that
\begin{equation}
\frac{1}{1+2c_0^2}(y'^2+z'^2)-x'^2=1\label{c_0}.
\end{equation}
That is, all the $\tilde{m}$ form a hyperboloid.

\end{proof}

\begin{thm} Let $\cal P$ be a nontrivial parity operator and let us consider the time reversal operators of the form ${\cal
T}=\displaystyle\sum_{i=0}^3 c_i\tau_i$ commuting with $\cal P$. All the points
$\tilde{c}=(c_1,c_2,c_3)$ form an ellipse. The length of the semi-major axis is
$\|\tilde{f}\|$ and the length of the semi-minor axis is $1$.
\end{thm}

\begin{proof}
By (\ref{c2}) and (\ref{c3}), we know that both $\tilde{b}$ and $\tilde{c}$ are orthogonal to $\tilde{f}$.


Construct a new $X'Y'Z'$ coordinate system by taking the direction of $\tilde{f}$ as that of the $Z'$ axis and the direction of $\tilde{b}$ as that of the $X'$ axis ( If $\tilde{b} = 0$, take any vector orthogonal to $\tilde{f}$ as the direction vector of the $X'$ axis ).
Then we have $\tilde{b}=(x,0,0)$, $\tilde{f}=(0,0,z)$ and $\tilde{c}=(c_1',c_2',c_3')$ in the $X'Y'Z'$ coordinate system.
Now the conditions
$(\ref{c1})-(\ref{c5})$ will reduce to
\ben
&&xc_3'=0,\label{d1}\\
&&xc_2'+c_0z=0,\label{d2}\\
&&zc_3'=0,\label{d5}\\
&&z^2-x^2=1,\label{d3}\\
&&(c_1')^2+(c_2')^2+(c_3')^2-(c_0)^2=1,\label{d4}.
\een
Note that (\ref{d3}) ensures that $z\neq 0$. Thus, (\ref{d1}) and (\ref{d5}) imply that $c_3'=0$, $\tilde{c}=(c_1',c_2',0)$.
In addition, it follows from (\ref{d2}) that $c_0=-\frac{x}{z}c_2'$.
Substituting $c_3'=0$, $c_0=-\frac{x}{z}c_2'$ and (\ref{d3}) into (\ref{d4}), we have
\be
(c_1')^2+\frac{(c_2')^2}{(z)^2}=1.
\ee

This is an equation of ellipse. Moreover, since $|z|=\|\tilde{f}\|>1$, the length of the semi-major axis is
$\|\tilde{f}\|$ and the length of the semi-minor axis is $1$.

\end{proof}

In the following theorem, we only consider the $\cal T$ with real coefficients.
\begin{thm} Let ${\cal T}_1$, ${\cal T}_2$ be two time reversal
operators, ${\cal T}_1\neq \pm{\cal T}_2$. If there exist two
nontrivial parity operators ${\cal P}_1$ and ${\cal P}_2$ such that ${\cal P}_i$ commutes with ${\cal T}_1$ and ${\cal T}_2$ simultaneously, then ${\cal
P}_1= \pm{\cal P}_2$.
\end{thm}

\begin{proof}  Let ${\cal T}_1=\displaystyle\sum_{i=0}^3 c_i^{(1)}\tau_i$, ${\cal T}_2=\displaystyle\sum_{i=0}^3 c_i^{(2)}\tau_i$. Denote
$\tilde{c}^{(1)}=(c_1^{(1)}, c_2^{(1)}, c_3^{(1)})$ and $\tilde{c}^{(2)}=(c_1^{(2)}, c_2^{(2)}, c_3^{(2)})$.

(i). If $c_0^{(1)}\neq0$ and $c_0^{(2)}=0$.

Suppose that $\cal P$ commute with ${\cal T}_i$ simultaneously.
 By (\ref{c1}), we have $\tilde{c}^{(2)}\times \tilde{b}=0$. It follows that
$\tilde{b}=m\tilde{c}^{(2)}$. On the other hand, (\ref{c1}) implies that $\tilde{f}=\frac{1}{c_0^{(1)}}(\tilde{c}^{(1)}\times \tilde{b})$.
Thus, $\tilde{f}=\frac{m}{c_0^{(1)}}(\tilde{c}^{(1)}\times \tilde{c}^{(2)})$. Substituting $\tilde{f}$ and $\tilde{b}$ into (\ref{c4}), then we
have
\[
m^2(\|\frac{1}{c_0^{(1)}}\tilde{c}^{(1)}\times \tilde{c}^{(2)}\|^2-\|\tilde{c}^{(2)}\|^2)=1.
\]
The equation has at most two real roots, which are opposite to each other. Thus, there exist at most two parity operators $\cal P$ and $-\cal P$ commuting with ${\cal T}_i$ simultaneously.

(ii). If $c_0^{(1)}=c_0^{(2)}=0$ and $\tilde{c}^{(1)}= t\tilde{c}^{(2)}$, where $t$ is a real
number.

It follows from (\ref{c5}) that ${\cal
T}_1 =\pm{\cal T}_2$, which contradicts with the assumption of the theorem.

(iii). If $c_0^{(1)}\neq0$, $c_0^{(2)}\neq0$ and $\tilde{c}^{(1)}= t\tilde{c}^{(2)}$.

By (\ref{c1}),
we have $\tilde{f}=\frac{1}{c_0^{(1)}}(\tilde{c}^{(1)}\times
\tilde{b})=\frac{1}{c_0^{(2)}}(\tilde{c}^{(2)}\times
\tilde{b})=\frac{t}{c_0^{(1)}}(\tilde{c}^{(2)}\times \tilde{b})$. It follows that $c_0^{(1)}=tc_0^{(2)}$. Thus, we have
 $c_i^{(1)}=tc_i^{(2)}$, $(i=0, 1, 2, 3)$.
On the other hand, (\ref{c5}) implies that $t^2=1$. Hence ${\cal T}_1=\pm {\cal T}_2$, which is a contradiction.

(iv). If $c_0^{(1)}=c_0^{(2)}=0$ and $\tilde{c}^{(1)}\neq t\tilde{c}^{(2)}$.

By (\ref{c1}), we have $\tilde{c}^{(1)}\times \tilde{b}=\tilde{c}^{(2)}\times \tilde{b}=0$. However, since $\tilde{c}^{(1)}\neq t\tilde{c}^{(2)}$, we have $\tilde{b}=0$. Thus (\ref{c4}) implies that $\|\tilde{f}\|=1$. Moreover, (\ref{c2}) implies that $\tilde{f}$ is orthogonal to both
$\tilde{c}^{(1)}$ and $\tilde{c}^{(2)}$. So $\tilde{f}$ can only have two directions, which are opposite to each other. Thus, there exist at most two parity operators $\cal P$ and $-\cal P$ commuting with ${\cal T}_i$ simultaneously.

(v). If $c_0^{(1)}\neq0$, $c_0^{(2)}\neq0$ and $\tilde{c}^{(1)}\neq t\tilde{c}^{(2)}$.

Let ${\cal P}_1$ and ${\cal P}_2$ be two parity operators, which are determined by $(\tilde{f}^{(1)},\tilde{b}^{(1)})$ and $(\tilde{f}^{(2)},\tilde{b}^{(2)})$ respectively. Moreover, suppose that both ${\cal P}_1$ and ${\cal P}_2$ commute with ${\cal T}_i$
simultaneously.

%

By (\ref{c1}), we have $\tilde{f}^{(1)}=\frac{1}{c_0^{(1)}}(\tilde{c}^{(1)}\times \tilde{b}^{(1)})$ and
$\tilde{f}^{(1)}=\frac{1}{c_0^{(2)}}(\tilde{c}^{(2)}\times \tilde{b}^{(1)})$. It follows that
\[
\frac{1}{c_0^{(1)}}\tilde{c}^{(1)}-\frac{1}{c_0^{(2)}}\tilde{c}^{(2)}=t_1\tilde{b}^{(1)},
\]
where $t_1$ is a nonzero real number.

Similarly, we have $\tilde{f}^{(2)}=\frac{1}{c_0^{(1)}}(\tilde{c}^{(1)}\times \tilde{b}^{(2)})$ and
$\tilde{f}^{(2)}=\frac{1}{c_0^{(2)}}(\tilde{c}^{(2)}\times \tilde{b}^{(2)})$. It follows that
\[
\frac{1}{c_0^{(1)}}\tilde{c}^{(1)}-\frac{1}{c_0^{(2)}}\tilde{c}^{(2)}=t_2\tilde{b}^{(2)},
\]
So $t_1\tilde{b}^{(1)}=t_2\tilde{b}^{(2)}$, which implies that $\tilde{b}^{(1)}=k\tilde{b}^{(2)}$. Now
$\|\tilde{f}^{(1)}\|^2-\|\tilde{b}^{(1)}\|^2=k^2(\|\tilde{f}^{(2)}\|^2-\|\tilde{b}^{(2)}\|^2)=1$,
hence $k=\pm1$. Thus it is apparent that ${\cal P}_1=\pm {\cal P}_2$.

Note that (i) -- (v) contain all the situations, which completes the proof.
\end{proof}

If we denote $com({\cal T})=\{{\cal P}|{\cal PT}={\cal TP}, {\cal
P}^2=I\}$, then the following corollary can be obtained.

\begin{cor} If ${\cal T}_1=\displaystyle\sum_{i=0}^3 c_i^{(1)}\tau_i$, ${\cal T}_2=\displaystyle\sum_{i=0}^3 c_i^{(2)}\tau_i$  are two time reversal operators, ${\cal T}_j^2=I$, $j=1, 2$. Then $com({\cal T}_1) = com({\cal T}_2)$ if and only if for each $i$, $c_i^{(1)}=\epsilon c_i^{(2)}$, where $\epsilon$ is a unimodular coefficient.
\end{cor}

\section{$\cal PT$-symmetric operators and unbroken $\cal PT$-symmetric condition}

A linear operator $H$ is said to be $\cal PT$-symmetric if $H{\cal
PT}={\cal PT}H$. As is known, in standard quantum mechanics, the
Hamiltonians are assumed to be Hermitian such that all the eigenvalues
are real and the evolution is unitary. In the $\cal
PT$-symmetric quantum theory, Bender replaced the Hermiticity of the
Hamiltonians with $\cal PT$-symmetry. However, the $\cal
PT$-symmetry of a linear operator does not imply that its eigenvalues must be real. Thus,
Bender introduced the unbroken $\cal PT$-symmetric condition. The
Hamiltonian $H$ is said to be unbroken $\cal PT$-symmetric if there
exists a collection of eigenvectors $\Psi_i$ of $H$ such that they
span the whole space and ${\cal PT}\Psi_i=\Psi_i$. It was shown
that for a $\cal PT$-symmetric Hamiltonian $H$, its eigenvalues are
all real if and only if $H$ is unbroken $\cal PT$-symmetric
\cite{bender2007making}. In two dimensional space case, this condition has a
much simpler description and an important illustrative example. That is, if
${\cal P}=\left(\begin{array}{*{2}{c@{\;\;}}c}
0& 1\\
1& 0\\
\end{array} \right)$, ${\cal T}={\cal T}_0$, $H=\left(\begin{array}{*{2}{c@{\;\;}}c}
re^{i\theta}& s\\
s& re^{-i\theta}\\
\end{array} \right)$, then $H$ is unbroken ${\cal PT}$-symmetric iff $s^2\geq r^2\sin^2\theta$ \cite{bender2007making}.

In the following part, we shall give the unbroken ${\cal PT}$-symmetry condition for general ${\cal PT}$-symmetric operators.
To this end, we need the following proposition.

\begin{prop} \label{prop} If $H$ is a $\cal PT$-symmetric operator, then it has four real parameters. Moreover, if  $H=h_0\sigma_0+h_1\sigma_1+
h_2\sigma_2+h_3\sigma_3$ is written in terms of Pauli operators,
then
\begin{eqnarray}
&&Im(h_0)=0\label{condition1},\\
&&Re(h_1)Im(h_1)+Re(h_2)Im(h_2)+Re(h_3)Im(h_3)=0.\label{condition2}
\end{eqnarray}
\end{prop}
\bpf
It is apparent that $\cal PT$ is also a time reversal operator. Thus we can assume that ${\cal PT}=\displaystyle\sum_{j=0}^3 c_j \tau_j$.
Now the condition ${\cal PT}H=H{\cal PT}$ is equivalent to
\bea
(\displaystyle\sum_{j=0}^3 c_j\tau_0\sigma_j)(\displaystyle\sum_{i=0}^3 h_i\sigma_i)=(\displaystyle\sum_{i=0}^3 h_i\sigma_i)(\displaystyle\sum_{j=0}^3 c_j\tau_0\sigma_j).
\eea

According to (\ref{pc}), this equation can be reduced to
\bea
c_0(\overline{h_0}-h_0)+\displaystyle\sum_{i=1}^3c_i(h_0-\overline{h_0})\sigma_i+\displaystyle\sum_{i=1}^3c_i(h_i+\overline{h_i})+
i\sigma\cdot[\tilde{c}\times(\overline{\tilde{h}}-\tilde{h})]-\displaystyle\sum_{i=1}^3c_0(h_i+\overline{h_i})\sigma_i=0,
\eea
where $\tilde{h}=(h_1, h_2, h_3)$ and $\overline{\tilde{h}}=(\overline{h_1}, \overline{h_2}, \overline{h_3})$.

The equation above is equivalent to
\ben
&&Im(h_0)=0,\label{53}\\
&&\displaystyle\sum_{i=1}^3 c_i Re(h_i)=0,\label{54}\\
&&\tilde{c}\times Im(h)-c_0Re(h)=0,\label{55}
\een
where $Re(h)=(Re(h_1),Re(h_2),Re(h_3))$ and $Im(h)=(Im(h_1), Im(h_2), Im(h_3))$.

(i). When $c_0\neq 0$.
 It follows (\ref{55}) that $Re(h)=\frac{1}{c_0}(\tilde{c}\times Im(h))$. Thus, the four parameters $Im(h)_1$, $Im(h)_2$, $Im(h)_3$ and $Re(h_0)$ determine $H$.

Note that (\ref{condition1}) is the same as (\ref{53}). On the other hand, $Re(h)=\frac{1}{c_0}(\tilde{c}\times Im(h))$ implies that $Re(h)\cdot Im(h)=0$. Thus, (\ref{condition2}) is also valid.

(ii). When $c_0=0$. (\ref{55}) implies that $Im(h)=t\tilde{c}$. Thus, we only need one real parameter $t$ to determine $Im(h)$. (\ref{54}) implies that $Re(h)$ should be orthogonal to $\tilde{c}$. Hence two parameters are needed. With $Re(h_0)$, we have four parameters altogether.

In this case, (\ref{condition2}) follows from the fact $Im(h)=t\tilde{c}$ and the equation (\ref{54}).

\epf

\begin{thm}\label{thm3} If $H$ is a $\cal PT$-symmetric operator and $\left(\begin{array}{*{2}{c@{\;\;}}c}
h_{11} & h_{12} \\
h_{21} & h_{22}\\
\end{array} \right)$ is the representation matrix of $H$, then $H$ is unbroken if and only if
$(Re(h_{11}+h_{22}))^2-4Re(h_{11}h_{22}-h_{12}h_{21})\geq 0.$
\end{thm}
\bpf
Let $\left(\begin{array}{*{2}{c@{\;\;}}c}
h_{11} & h_{12} \\
h_{21} & h_{22}\\
\end{array} \right)$ be the matrix of $H$, $\lambda$ be an eigenvalue of $H$, then
\be
\lambda^2-(h_{11}+h_{22})\lambda+h_{11}h_{22}-h_{12}h_{21}=0.\label{delta}
\ee

On the other hand, rewrite $H=h_0\sigma_0+h_1\sigma_1+h_2\sigma_2+h_3\sigma_3$.
It follows from (\ref{condition1}) and (\ref{condition2}) that
\begin{eqnarray*}
&&Im(h_{11}+h_{22})=2Im(h_0)=0,\\
&&Im(h_{11}h_{22}-h_{12}h_{21})=-Re(h_1)Im(h_1)-Re(h_2)Im(h_2)-Re(h_3)Im(h_3)=0.
\end{eqnarray*}
The two equations above imply that
\be
-Im(h_{11}+h_{22})\lambda+Im(h_{11}h_{22}-h_{12}h_{21})=0.\label{u2}
\ee

Substitute (\ref{u2}) into (\ref{delta}). Now the equation (\ref{delta}) reduces to
\be
\lambda^2-Re(h_{11}+h_{22})\lambda+Re(h_{11}h_{22}-h_{12}h_{21})=0,\label{u1}
\ee
According to (\ref{u1}), $\lambda$ is a real number, that is, $H$ is unbroken
$\cal PT$-symmetric, if and only if
\begin{equation}
(Re(h_{11}+h_{22}))^2-4Re(h_{11}h_{22}-h_{12}h_{21})\geq 0.
\label{eq::u1}
\end{equation} \epf

\begin{remark}
Note that when the equality is valid in (\ref{eq::u1}), $H$ may be non-diagonalisable in general. In this case, the space $\mathbb C^2$ is actually spanned an eigenvector $\psi_1$ satisfying $(H-\lambda_0 I)\psi_1=0$ and a generalized eigenvector $\psi_2$ satisfying $(H-\lambda_0 I)^2\psi_2=0$, where $\lambda_0=\frac{1}{2}Re(h_{11}+h_{22})$ is the eigenvalue.
\end{remark}

\begin{remark} Note that Bender's unbroken $\cal PT$-symmetric condition in \cite{bender2007making} is a special case of (\ref{eq::u1}). To see this, let
$H=\left(\begin{array}{*{2}{c@{\;\;}}c}
re^{i\theta}& s\\
s& re^{-i\theta}\\
\end{array} \right),$ we have $$Re(h_{11})=Re(h_{22})=r\cos\theta,$$ $$Re(h_{11}h_{22}-h_{12}h_{21})=r^2-s^2.$$ Then  (\ref{eq::u1}) holds iff  $s^2\geq r^2\sin^2\theta$.
\end{remark}

\begin{remark}\label{thm2} If $H$ is a Hermitian operator, then it is also unbroken $\cal PT$-symmetric. Usually, this can be shown by using canonical forms. However, in $\mathbb C^2$, it also follows from direct calculation.
\end{remark}
In fact, since $H=h_0\sigma_0+h_1\sigma_1+
h_2\sigma_2+h_3\sigma_3$ is Hermitian, each $h_i$ is a real number. Now
we only need to find real coefficients $c_0$, $c_1$, $c_2$ and $c_3$ such that $c_1^2+c_2^2+c_3^2-c_0^2=1$ and
equations $(\ref{53})-(\ref{55})$ hold. Take $c_0=0$ and $c_1$, $c_2$, $c_3$ are such real numbers that $c_1Re(h_1)+c_2Re(h_2)+ c_3Re(h_3)=0$ and
$c_1^2+c_2^2+c_3^2=1$. Let ${\cal PT}=\displaystyle \sum_{i=0}^3 c_i\tau_i$. It is apparent that $({\cal PT})^2=I$ and $H$ is
$\cal PT$-symmetric. Moreover, if we rewrite the Hermitian matrix as $H=\begin{pmatrix}
a&b\\
\overline{b}&a\end{pmatrix}$,
then
$Re(h_{11}+h_{22})^2-4Re(h_{11}h_{22}-h_{12}h_{21})=4a^2-4(a^2-|b|^2)=4|b|^2\geqslant
0$ holds, so $H$ is also unbroken.

\end{document}